\documentstyle[preprint,floats,prc,aps]{revtex}
\begin{document}

% uncomment \draft to have PACS numbers appear
\draft

% put preprint number(s).
\preprint{IU/NTC 94-18 //OSU--94-147}

\tighten

\title{Vacuum Contributions in a Chiral Effective\\ Lagrangian
 for  Nuclei}

\author{R. J. Furnstahl and Hua-Bin Tang}
\address{Department of Physics\\
         The Ohio State University,\ \ Columbus,~Ohio\ \ 43210}
\author{Brian D. Serot}
\address{Physics Department and Nuclear Theory Center\\
         Indiana University,\ \ Bloomington,~Indiana\ \ 47405}
%\date{DRAFT: \today}
\date{January 25, 1995}
\maketitle

\begin{abstract}
A relativistic hadronic model for nuclear matter and finite nuclei,
which incorporates nonlinear chiral symmetry and
broken scale invariance, is presented and applied at the one-baryon-loop
level to finite nuclei.
The model contains an effective light scalar field that is responsible
for the mid-range nucleon--nucleon attraction and which has anomalous
scaling behavior.
One-loop vacuum contributions in this background scalar field
at finite density
are constrained by  low-energy theorems that reflect the broken
scale invariance
of quantum chromodynamics.
A mean-field energy functional for nuclear matter and nuclei is derived
that contains small powers of the fields and their derivatives, and the
validity
of this truncation is discussed.
Good fits to the bulk properties of finite nuclei and single-particle
spectra are obtained.
\end{abstract}

\vspace{20pt}
%
%\pacs{PACS number(s): 21.30.+y, 21.60.Jz, 21.65.+f}
%
\def\capcrunch{%
%    \protect\small%                                  change font
    \setlength{\baselineskip}{0.9\baselineskip}%     change leading
    \setlength{\hangindent}{0.04\textwidth}%         indent
    \setlength{\hsize}{0.94\textwidth}%              change right
                                     }%              margin
\narrowtext

\section{Introduction}

Descriptions of nuclear matter and finite nuclei, which are ultimately
governed by the physics of low-energy quantum chromodynamics (QCD),
are efficiently formulated using low-energy degrees of
freedom---the hadrons.
In the absence of direct derivations from QCD, such effective descriptions
should be constrained by the underlying
symmetries of QCD, both broken and unbroken.
Nevertheless, the appropriate realization of these symmetries for
phenomenological models is not yet established.
In this paper, we explore some consequences of applying QCD symmetry
constraints to a relativistic model of finite nuclei that features a
light scalar meson.

At present,
the most developed framework for constraining hadronic physics by
QCD symmetries is
chiral perturbation theory (ChPT)\cite{WEINBERG68},
which provides a systematic expansion in
energy for low-energy scattering processes.
The degrees of freedom are the Goldstone bosons (pions, {\it etc}.)
and, when appropriate, nucleons.
This approach builds in constraints due to chiral symmetry without any
additional constraints on the dynamics or {\it ad hoc\/} model assumptions;
physics beyond chiral symmetry is incorporated through constants
in the low-energy lagrangian,
which are usually determined from experiment.
Because additional constants are needed at each stage in the energy
expansion,
ChPT is predictive only
at sufficiently low energies, where the number
of parameters introduced does not overwhelm the data to be described.

The prospects for extending ChPT in a useful way to calculations at
finite density are unclear at present.
On the other hand, the general framework of ChPT has validated the
principle of resonance dominance of low-energy QCD.
In particular, the $E^4$ coupling constants in ChPT in the meson sector
are well reproduced
from a meson resonance lagrangian applied at tree level,
with the vector mesons playing the leading role\cite{ECKER89}.
Meson dominance is also the key principle
underlying phenomenological models of nuclei with hadronic degrees of freedom,
which we consider here.
But while the correspondence
in the vector channels is relatively straightforward
because of well-defined resonances, the dynamics in the
scalar channel is more difficult to identify and to model.

Within meson-exchange phenomenology, the mid-range attraction between nucleons
is generally believed to be a dynamical
consequence of the strong interactions between two pions
exchanged with scalar, isoscalar quantum numbers \cite{DURSO}.
No nearby underlying resonance at the relevant mass ($\approx 500\,$MeV)
is evident or, in principle, needed.
(Note that in ChPT investigations, the scalar resonance is identified
with mesons around 1~GeV.)
Nevertheless, this physics is efficiently, conveniently, and adequately
represented at the one-meson exchange
level by the exchange of a light scalar degree of freedom\cite{MACHLEIDT}.
This light scalar is also an essential element of phenomenologically
successful mean-field models of nuclei\cite{WALECKA74,SW}.

These mean-field models are significantly constrained by the bulk properties
of finite nuclei\cite{REINHARD89,BODMER91,FS}.
The question then arises:
How should QCD symmetry constraints be manifested in these models?
There is a long history of
attempts to generalize the linear sigma model to build models with
chiral symmetry;
it is almost irresistible to identify the scalar meson mediating the
mid-range nucleon--nucleon (NN) attraction with the chiral partner of the pion.
More recently, interest in models realizing the broken scale invariance
of QCD has been revived.
Scale invariance is particularly compelling to consider because of its
connection to the scalar channel.
The breaking of scale invariance by the trace anomaly
implies relations involving
zero-momentum Green's functions
of the scalar trace of the energy-momentum tensor
[see Eqs.~(\ref{eq:leta})--(\ref{eq:letb})];
these are called low-energy theorems\cite{VAINSHTEIN82}.
If these relations are assumed to be
saturated by scalar particles at tree level,
significant constraints arise on
the associated scalar potentials (in the chiral limit).
We will exploit such constraints in this paper.

In Ref.\cite{FS}, a broad class of models that attempt to
unite successful mean-field phenomenology with chiral symmetry and
the broken scale invariance of QCD were studied.
Generalizations of the conventional linear sigma model that feature
a ``Mexican hat'' potential were found to fail generically,
even with  modifications inspired by the realization of broken scale
invariance.
A significant improvement was found by the Minnesota group\cite{HEIDE94}
when the ``Mexican hat'' potential is  abandoned, and a reasonable
description of the properties of closed-shell nuclei was obtained.

In this paper, we build a different effective model of nuclei
by implementing a {\em nonlinear\/} realization of chiral symmetry together
with the low-energy theorems of broken scale invariance.
The detailed construction
of the full model will be reported elsewhere\cite{FST}.
Our focus here is primarily on how vacuum dynamics might be treated in an
effective field theory of nuclei.

The role and manifestation of vacuum dynamics is
an important issue in any field-theoretic description of
nuclear matter and finite nuclei.
Valence nucleons in the Fermi sea interact with each other and also
with the QCD vacuum.
In turn, the vacuum is modified by interactions with valence nucleons.
In nonrelativistic models, such effects are never dealt with explicitly, but
are absorbed implicitly
into phenomenological effective interactions involving only valence nucleons.
As a result, the interactions may acquire
additional density dependence and nonlocalities.
In previous
relativistic models of nuclear matter involving a scalar field coupled
to the nucleon, vacuum modifications were incorporated in the
renormalized scalar effective potential \cite{WALECKA74,SW}.
This in turn affects the density dependence.

In principle, the one-baryon-loop
effective potential contains an infinite number
of undetermined coupling constants, which are the coefficients in
a polynomial of infinite order in the scalar field.
In conventional renormalizable models, the
nucleon vacuum one-loop correction is well defined\cite{CHIN77} and
determines these coefficients, except for the terms of degree four and less,
which are fixed by a renormalization prescription.
However,
renormalizable models with one-loop corrections do not achieve the
phenomenological success of models without vacuum terms for the bulk
properties of finite nuclei \cite{FOX,FPW}.
We interpret this failure as a phenomenological indication that the
vacuum is not treated adequately.

In previous studies involving nonrenormalizable models,
the effective potential is simply truncated, usually at degree four,
and mean-field theory is applied without considering
vacuum effects.
In this paper, we begin to address the problem of constructing
consistent calculations
in effective field theories
of nuclear matter and finite nuclei that
explicitly address the role of the vacuum dynamics.
In particular,
we show how vacuum loop contributions are absorbed in   the
renormalization of coupling constants in the lagrangian
in a model constrained to satisfy the low-energy theorems of QCD.

In contrast to the situation in ChPT, we cannot
expand in powers of the energy,
since we are not limited to derivative couplings and light meson
masses.
We observe, however, that the meson fields develop nonzero expectation values
(mean fields) at finite density, and to begin, we assume that these mean
fields dominate the contributions to the energy.
Successful mean-field phenomenology shows that for densities not much higher
than nuclear matter equilibrium density, the corresponding mean fields (or
nucleon self-energies) are small compared to the free nucleon mass
(roughly $\textstyle{1 \over 4}$ to $\textstyle{1 \over 3}$ the size); these
ratios are therefore useful expansion parameters.
Moreover, since the derivatives of the mean fields are small for normal
nuclei, a truncation of the lagrangian at some low order of derivatives is
also appropriate.
(We verify this assertion explicitly later.)
The end result is an energy functional for nuclear matter and nuclei that
contains small powers of the mean fields and their derivatives; nuclear
phenomenology implies that these fields are an efficient way to incorporate
the density dependence of nuclear observables.
Our objective here is to see how the low-energy behavior of QCD constrains
the coefficients in this energy functional, particularly with regard to
contributions from the quantum vacuum.

The assumption of mean-field dominance also has phenomenological support
from Dirac--Brueckner--Hartree--Fock (DBHF)
calculations, which indicate that exchange terms and
short-range correlations do not significantly
change the size of the nucleon self-energies nor
introduce strong momentum dependence (at least for occupied states)
\cite{CJHBDS,TERHAAR,MACHLEIDT}.
Thus we have the favorable situation that the mean fields are large enough
(compared to nuclear energy scales)
to dominate the bulk dynamics
but small enough (compared to the nucleon mass) to provide useful
expansion parameters.
Going beyond one-loop order systematically is an essential issue, but we will
leave this as a topic for future study.

A nonlinear realization of chiral symmetry will be adopted, in which the
Goldstone bosons (pions) are derivatively coupled to the nucleons.
Historically,
a linear representation (as in the usual linear sigma model) has been favored
by model builders, in part because the sigma model is renormalizable.
In this work, we wish to introduce a light scalar degree of freedom, but
we do not want to make the restrictive dynamical assumption that this scalar is
the chiral partner of the pions in a linear representation.
By realizing chiral symmetry nonlinearly, we are not committed to
such assumptions about the scalar degree of freedom.
In addition, it will be easier to introduce vector mesons
in a chirally invariant way that manifests the
vector-meson dominance of Sakurai\cite{SAKURAI69}.
Finally, the nonlinear representation is more efficient for preserving
the consequences of chiral symmetry at finite density when making
approximations involving pions, because sensitive cancellations are not
needed\cite{SW}.%
\footnote{We are unaware, however, of any proof
of the independence of finite-density observables
with respect to nonlinear field transformations,
analogous to the theorem that applies to $S$-matrix elements \cite{SMATRIX}.}

As suggested above,
the nonderivative terms of the  light scalar effective potential
can be constrained by the
low-energy theorems of QCD, so that vacuum effects are
``built in.''
This, together with the truncation of our expansion
in derivatives and powers of fields, leaves us with relatively few
parameters, which can be determined by fitting to the properties of
finite nuclei.

The paper is organized as follows:
In Section II, broken scale invariance is discussed and
the model is introduced and its renormalization is considered.
An approximation scheme for nuclear matter and finite nuclei
is proposed in Section III and the energy functional is derived.
Results are given in Section IV.
Section V contains some discussion of the results and Section VI is
a summary.

%%%%%%%%%%%%%%%%%%%%%%%%%%%%%%%%%%%%%%%%%%%%%%%%%%%%%%%%%%%%%%%%%%%%%%%
%%%%%%%%%%%%%%%%%%%%%%%%%%%%%%%%%%%%%%%%%%%%%%%%%%%%%%%%%%%%%%%%%%%%%%%

\section{The Model}

In meson-exchange phenomenology there is a light scalar degree of freedom
that simulates two-pion-exchange physics in the scalar channel
\cite{DURSO,LIN}.
Here we seek to describe this physics by introducing a
light scalar field $S(x)$.
We do not associate the scalar with a bound state or resonance,
so we allow $S(x)$ to have anomalous behavior under a scale transformation
in the effective theory.
In particular, when $x\rightarrow \lambda^{-1} x$,
$S(x)\rightarrow \lambda^d S(\lambda x)$, where $d$ can differ from unity
and is to be determined phenomenologically.
A QCD-inspired scenario that leads to such a scalar was proposed by
Miransky and collaborators\cite{MIRANSKY}.
They introduced a light scalar generated by
dynamical chiral symmetry breaking in QCD, which was consequently
associated with the
quark condensate $\langle\bar{q}q\rangle$ and referred
to as quarkonium.
We will take all other fields to have canonical scale dimension.

While massless QCD is scale invariant at the classical level, this symmetry
is broken at the quantum level.
This breaking is manifested in a nonzero trace of the energy-momentum
tensor of QCD, which is referred to as the trace anomaly.
The QCD trace anomaly in the chiral limit\cite{COLLINS77}
is given by
\begin{equation}
 \theta_{\mu}^{\mu}(x)\equiv-H(x)=(\beta(g)/2g)
G_{\mu\nu}^aG^{a\mu\nu} \ , \qquad a=1, 2, \cdots, N_{\rm c}^2-1
                 \ ,
\end{equation}
where $G^{a}_{\mu\nu}$ is the gluon field tensor and
$\beta(g)=-(g^3/48\pi^2)(11N_{\rm c}-2N_{\rm f})$ is the one-loop beta
function with $N_{\rm c}$ colors and ${N_{\rm f}}$ flavors.
There are remnants of scale invariance, which imply low-energy theorems that
relate Green's functions involving the trace of the
energy-momentum tensor $H(x)$\cite{VAINSHTEIN82}:
\begin{eqnarray}
i\int{\rm d}^4x\, \langle 0|T[H(x)H(0)]|0\rangle  & = &
4H_{\scriptscriptstyle 0} \ , \label{eq:leta} \\[4pt]
i^2\int{\rm d}^4x \, {\rm d}^4y \, \langle 0|T[H(x)H(y)H(0)]|0\rangle
    & = & 4^2H_{\scriptscriptstyle 0} \ , \label{eq:letb} \\[4pt]
   \vdots  \qquad\qquad & & \nonumber
\end{eqnarray}
where $H_{\scriptscriptstyle 0}\equiv \langle 0|H|0\rangle$.

Effective lagrangians for pure-glue QCD (no quarks) featuring a scalar
glueball field $\chi(x)$
(``gluonium'') that saturates these low-energy theorems at tree level
have been considered many times \cite{SCHMIGELL}.
Lattice QCD calculations indicate that the scalar glueball is quite
heavy on hadronic mass scales, with a mass of roughly
1.6--1.8\,GeV \cite{LATTICE}.
Its fate in the real world with light quarks is not entirely clear.
Here we will generalize the effective gluonium
model to include the light scalar discussed earlier;
this extension was proposed in a different context in Ref.\cite{MIRANSKY}.
We take the trace anomaly to consist of two contributions,
corresponding to a  vacuum expectation value
$H_{\scriptscriptstyle 0}=H_{\rm g}+H_{\rm q}$.
Here $H_{\rm g}$  is identified with the
heavy glueball contribution, while $H_{\rm q}$ is nonzero only
when chiral symmetry is dynamically broken in the presence of light quarks.
One can argue that $H_{\rm g}$
dominates $H_{\scriptscriptstyle 0}$ (which is equal to the gluon condensate
up to a factor) so that $H_{\rm q} \ll H_{\rm g}$ \cite{MIRANSKY}.
How the QCD trace anomaly actually separates into the two parts is not
explored here, since we will determine $H_{\rm q}$ by fitting to the
properties of  finite nuclei.
Nevertheless, we find that the value of $H_{\rm q}$ determined
in our fits  satisfies $H_{\rm q} \ll H_0$ (see Table~\ref{tab:one}).

The low-energy theorems involving  the trace
$\theta_{\mu}^{\mu}(x)$ of the energy-momentum tensor are
assumed to be saturated by the scalar gluonium $\chi(x)$
\cite{SCHMIGELL} {\it and\/} the light scalar  $S(x)$.
For simplicity we adopt a model with no mixing between the scalars.
%We also assume
%the mixing between them to be negligible\cite{LATTICE}.
A candidate effective lagrangian
of the scalars that satisfies the low-energy theorems at
tree level in the chiral limit
is\cite{SCHMIGELL,MIRANSKY}
\begin{equation}
  {\cal L}_{\rm s}(x)={1\over 2}\partial _{\mu}\chi\partial
        ^{\mu}\chi
   + {1\over 2}\bigg [\alpha_1
   \bigg ({\chi^2 \over \chi_{ 0}^2}\bigg ) ^{1-d}
       +(1-\alpha_1)  \bigg (
       {S^2 \over S_{ 0}^2}\bigg )^{(1-d)/d}\
         \bigg ]
   \partial _{\mu}S\partial ^{\mu}S- V(\chi, S) \ , \label{eq:Lscalar}
\end{equation}
where $\alpha_1$ is a real constant, $d$ is the scale dimension of
the $S(x)$ field, and the scale-breaking potential $V$ is
\begin{equation}
V(\chi, S)  =    H_{\rm g}
          {\chi^4 \over \chi_{ 0}^4}
         \bigg ( \ln {\chi \over \chi_0} - {1\over 4}\bigg)
         +H_{\rm q}\bigg ({S^2 \over S_{ 0}^2}
         \bigg)^{2 /d}    \bigg ( {1 \over 2d}
              \ln {S^2 \over S_{ 0}^2}
                   - {1 \over 4} \bigg )   \ . \label{eq:potl}
\end{equation}
Here $ \chi_{ 0}$  and $S_{ 0}$
are  the vacuum expectation values of $ \chi$ and $S$ respectively.
Notice that $\alpha_1$ has been introduced so that after expanding the terms
in square brackets in Eq.~(\ref{eq:Lscalar}), the
kinetic term for $S$ is canonical.
The mass of the light scalar $S$ is given by
$m_{\rm s}^2 =  4H_{\rm q}/(d^2S_{ 0}^2)$.
The scale dimension of the $\chi$ field is assumed to be unity.

One can define the energy-momentum tensor so that the Noether current
for scale transformations is $x_\nu \theta^{\mu\nu}$.
The trace of this ``improved'' energy-momentum  tensor\cite{CALLAN70}
corresponding to the lagrangian in Eqs.~(\ref{eq:Lscalar})
and (\ref{eq:potl}) is
\begin{eqnarray}
\theta_{\mu}^{\mu}(x)& =& Sd{\partial V \over \partial S}
                +\chi{\partial V \over \partial \chi}
                    -4 V(\chi, S)      \nonumber \\
  & = &- H_{\rm g}
          {\chi ^4 \over \chi_{ 0}^4}
        -H_{\rm q}\biggl({S^2 \over S_{ 0}^2}
         \biggr)^{{2/ d}} \ . \label{eq:trace}
\end{eqnarray}
With the
dynamics of the scalar field fluctuations governed by the lagrangian
in Eqs.~(\ref{eq:Lscalar}) and (\ref{eq:potl}), the preceding trace
satisfies the low-energy theorems at the tree level.
The usual direct demonstration\cite{SCHMIGELL}, in which the gluonium
alone is assumed to saturate the Green's functions,
involves parametrizing
the fluctuations $\widetilde\chi(x)$ in the exponential form
$\chi=\chi_{ 0}\exp [\widetilde\chi(x)/
\chi_{ 0}]$ and substituting this into
the low-energy theorems and into the potential,
$V(\chi,0)$ of Eq.~(\ref{eq:potl}), to determine vertices.
Keeping only tree level diagrams (no loops), the theorems then follow.

To extend the demonstration to the present case,
one first notes that the low-energy theorems should not depend on
how the gluonium field fluctuation $\widetilde\chi$ is parametrized.
One then observes that if
$\widetilde {\chi}$ is defined through $\chi \equiv \chi_{ 0}
(1 - \widetilde {\chi}/\chi_{ 0})^d$, the
resulting form for the gluonium parts of the trace
and the potential in Eqs.~(\ref{eq:potl}) and (\ref{eq:trace})
become the same as for the light scalar,
when the fluctuation of the latter
is parametrized simply as $S = S_0 - \phi$.
Since there are no couplings between the $\chi$ and the $S$
fields, the low-energy theorems follow directly.

We can now add to ${\cal L}_s$
a {\it scale-invariant\/} lagrangian with these scalars coupled
to pion, nucleon, and
vector degrees of freedom.
(We neglect pion mass terms at this point.)
The resulting model would be a candidate model for nuclei that satisfies
the low-energy theorems.
On the other hand,
there are many other possible terms allowed,
and even Eq.~(\ref{eq:Lscalar}) is not
the most general form involving two scalars.
We choose to take advantage of the heaviness of the gluonium and
the usefulness of an expansion in powers and derivatives of the other fields
to both simplify and generalize the effective lagrangian.
We expect the expansion to be valid and
useful when applied near normal nuclear matter densities.

Since the mass scale of the heavy gluonium field (roughly 1.6--1.8\,GeV)
is significantly higher
than the scales involved in the nuclear matter problem,
the heavy gluonium field fluctuations $\widetilde\chi$
can be integrated out as in Ref.~\cite{ECKER89}.
In particular, we can eliminate $\widetilde\chi$ by iteratively solving
its equation of motion, exploiting the dominance of the mass term over
powers and derivatives of $\widetilde\chi$.
This results in complicated terms involving powers and derivatives
of the other fields, but we can expand these terms.
For example, the second term in Eq.~(\ref{eq:Lscalar}) would become
\begin{equation}
   {1\over 2}\biggl[1 + \beta_1 {\phi\over S_0} +
        \beta_2 {\phi^2\over S_0^2}
        + \cdots \biggr]
   \partial _{\mu}\phi\partial ^{\mu}\phi
   + \beta_3(\partial _{\mu}\phi\partial ^{\mu}\phi)^2 + \cdots ,
\end{equation}
where the $\beta_i$ are functions of the constants
$\chi_0$, $\alpha_1$, and $d$.
The lagrangian will be much simpler if we can
truncate this expansion at leading order in  derivatives and
neglect high powers of the meson fields.

We can follow this prescription to write a general chirally invariant
effective lagrangian for nuclear matter and nuclei.
The ground states of even-even nuclei and nuclear matter will be
assumed to have good parity, so there is no pion mean field.
Thus the pion will not play an explicit role in the present
discussion of uniform nuclear matter and closed-shell nuclei
in the Hartree approximation.
Nevertheless, we wish to stress the connection to pion physics and
the underlying constraints of chiral symmetry.
Thus,
we give an overview of the full model in order to motivate the form of
the lagrangian and to set the stage for future work.

We restrict consideration to a low-energy representation of
massless, two-flavor QCD.
The Goldstone pion fields are represented by a chiral phase
angle that corresponds to a pure chiral rotation of the identity matrix:
\begin{equation}
 \xi {\bf 1} \xi \equiv U(x) =
     \exp (i \bbox{\pi}(x) \bbox{\cdot \tau}/ f_{\pi})
               \ , \label{eq:Udef}
\end{equation}
where
$\xi (x) = \exp (i \bbox{\pi}(x)
\bbox{\cdot \tau}/ 2f_{\pi})$,
$\tau ^a$  ( $a=1, 2$, and $3$ ) are the Pauli matrices,
$\pi^a(x)$  are the Goldstone pion fields, and $f_{\pi}=93$ MeV
is the pion-decay constant.
This parametrization and the nucleon representation that follows is
conventional;
see, for example, Ref.~\cite{DONOGHUE92}.
The nucleon field is written as
\begin{equation}
N(x)=\left(\begin{array}{c} p(x) \\ n(x) \end{array} \right)\ ,
\end{equation}
with $p(x)$ and $n(x)$ being the proton and neutron fields.

Under  chiral transformations of
$SU(2)_{\rm L} \otimes SU(2)_{\rm R}$,
$U(x)$ transforms globally,  $U(x)\rightarrow L U(x) R^{\dagger}$,
where $L = \exp(i\bbox{\theta_L\cdot\tau})$ and
$R = \exp(i\bbox{\theta_R\cdot\tau})$
are $x$-independent elements of $SU(2)_{\rm L}$ and $SU(2)_{\rm R}$,
respectively.
In general, the transformation of $\xi$ is local since
it depends on the pion field \cite{GEORGI,DONOGHUE92,CALLAN69}:
\begin{equation}
\xi(x) \rightarrow \xi'(x) = L \xi(x) h^{\dagger}(x) = h(x) \xi(x) R^{\dagger}
               \ , \label{eq:Xitrans}
\end{equation}
where the second equation defines the $SU(2)$-valued function
$h(x)$ as a {\it nonlinear\/} function of $L$, $R$, and $U(x)$.
Note that $h=L$ when $L=R$, {\it i.e.,} in the case of a pure isospin
rotation.
The nucleon field also transforms locally:
$N(x)\rightarrow h(x) N(x)$,
which implies that nucleons mix with pions under chiral transformations.
(See Ref.~\cite{GEORGI} for alternative representations of the nucleon field.)

We will incorporate the physics of vector dominance in our lagrangian
by introducing vector mesons as gauge bosons \cite{SMATRIX}.
For simplicity, since
we concentrate on the properties of nearly symmetric
($N \approx Z$) nuclear matter in this
paper, we will not explicitly write down the rho and the
electromagnetic fields.
Thus only the $\omega$ meson field $V^\mu$ appears explicitly here.
We will present a full discussion of the lagrangian elsewhere
\cite{FST}, including
how the vector mesons are gauged and how vector dominance results.
To build chirally invariant terms,
it is useful to define the vector and axial vector fields
\begin{eqnarray}
v_{\mu}(x)  & = & -{i \over 2}(\xi^{\dagger} \partial_{\mu} \xi +
    \xi \partial_{\mu}\xi^{\dagger} )
            \ , \label{eq:vdef}  \\[4pt]
a_{\mu}(x)  & = & -{i \over 2}(\xi^{\dagger} \partial_{\mu} \xi -
    \xi \partial_{\mu}\xi^{\dagger} )  \  , \label{eq:adef}
\end{eqnarray}
which transform as
$v_{\mu}   \rightarrow          h v_{\mu} h^{\dagger}
                   -ih\partial_{\mu}h^{\dagger}$ and
$a_{\mu} \rightarrow h a_{\mu} h^{\dagger}$.
The coupling of the pion to the nucleon is realized through $a_\mu$
and the
covariant derivative
\begin{equation}
{\cal D}_{\mu}  =  \partial _{\mu}
                + i v_{\mu} +i g_{\rm v}V_{\mu}
           \ . \label{eq:nulcov}
\end{equation}

Now we can write the complete chirally invariant lagrangian;
all terms not contained in ${\cal L}_s$ are scale invariant.
After integrating out $\widetilde\chi$ and expanding about $S_0$,
the lagrangian takes the form
\begin{eqnarray}
{\cal L}(x) &=&
         \overline N \Bigl(i\gamma^{\mu} {\cal D}_{\mu}
             -ig_{\rm \scriptscriptstyle A}\gamma^{\mu}\gamma_5
             a_{\mu} - M  + g_{\rm s}\phi + \cdots \Bigr)N
       -{1\over 4}F_{\mu\nu}F^{\mu\nu}
                              \nonumber \\
 &  &  \null  + {1\over 2} \bigg [
      1+ \eta {\phi\over S_0}  + \cdots \bigg ]
         \Big [ {1\over 2}f_{\pi}^2\, {\rm tr}\,
                  (\partial _{\mu}U\partial ^{\mu}U^{\dagger})
                + m_{\rm v}^2 V_{\mu}V^{\mu} \Big ]
                                       \nonumber \\
         &  &       \null    +{1\over 4!}\zeta
             (g_{\rm v}^2 V_{\mu}V^{\mu})^2
             + {1\over 2}\partial_\mu \phi \partial^\mu \phi
         - H_{\rm q}\bigg ({S^2 \over S_{ 0}^2}
         \bigg)^{2 / d}    \bigg ( {1 \over 2d}
              \ln {S^2 \over S_{ 0}^2}
                   -{1 \over 4} \bigg )      + \cdots
                \ ,\label{eq:NLag}
\end{eqnarray}
where  $g_{\rm \scriptscriptstyle A}=1.23$ is the
axial coupling
constant, $g_{\rm s}$ ($g_{\rm v}$) is the light scalar
(vector $\omega$) coupling
to the nucleon, the $\omega$ field strength tensor
is $F_{\mu\nu}= \partial_{\mu} V_{\nu}
              -\partial_{\nu} V_{\mu}$,
and $\eta$ and $\zeta$ are real constants.

Several features of this lagrangian are of interest:
\begin{itemize}
  \item
  We have combined terms after expanding and have rewritten the
  coefficients,
  where appropriate, in terms of physical masses.
  Note that  the nucleon mass $M$ has contributions   from the vacuum
expectation values of both  scalars; we do not assume that it comes
entirely from the light scalar (although this possibility is not excluded).
  \item
 The combination of the $\omega$ mass term and the  pion kinetic
term in Eq.~(\ref{eq:NLag}) appears naturally, if we assume the
vector mesons to be  gauge bosons\cite{SMATRIX,FST}.
  \item
  The original separation of the lagrangian into a scale-invariant
  piece and a scale-breaking piece, in which  the latter
  involved only the scalar
  fields, is now largely hidden
  because the $\chi$ dependence is not explicit and we have expanded
  about $S_0$.
Nevertheless, there is a remnant for our purposes here: the
scale-breaking potential of the light scalar
[the last term in Eq.~(\ref{eq:NLag})], which is not changed by
the elimination of $\widetilde\chi$. (Recall that
  the $\chi$ and $S$ do not mix.)
Thus the low-energy theorems still protect the form
of this potential, which places  constraints on vacuum
loop renormalizations, as discussed below.
  \item
We have omitted many higher-order terms, as indicated
by the ellipses, which represent higher powers of fields and their
derivatives.
Only Yukawa couplings to the nucleon fields are kept, based on the
phenomenological dominance of one-meson exchange and the implicit
elimination of heavier fields.
(So $\overline NN\phi^2$ terms, {\it etc.}, are omitted.)
Higher-order terms with meson fields
should give numerically small contributions (in nuclei)
or can be absorbed into slight adjustments of the other parameters.
Some explicit justification for these claims is given in the results below.
\end{itemize}

The lagrangian in Eq.~(\ref{eq:NLag})
is written with renormalized coefficients.
Counterterms are not written explicitly, but are implied.
In particular,
these counterterms include {\it all\/} powers of the scalar field,
not just terms up to $O(\phi^4)$, as in a renormalizable model.
To understand how these counterterms are fixed, we start by
integrating out the baryon fields  at zero density and temperature.
The result is a fermion determinant that contributes to the
meson action as an additive term given by
\begin{equation}
S_{\text{fd}}[\phi,V_\mu] \equiv \int {\rm d}^4 x\, {\cal L}_{\text
{fd}}
                = -i\, {\rm Tr}\ln K(0)
           \ , \label{eq:Seff}
\end{equation}
where
``Tr'' indicates a trace over spacetime, spin, and isospin, and
the kernel $K(\mu)$ is defined in coordinate space by
\begin{equation}
  \langle x | K(\mu) | y \rangle =
   [i\gamma^\mu \partial_\mu  - g_{\rm
v}\gamma^{\mu}V_{\mu}(x)
               +\mu \gamma_0     -M+g_{\rm s}\phi(x)] \delta^4(x-y) \ .
\end{equation}
The introduction of the chemical potential $\mu$ is for later
convenience, and baryon counterterms, which are needed beyond one-loop,
are suppressed.
Note that no approximation has been made at this point;
$S_{\text{fd}}$ is a functional of the dynamical fields $\phi$
and $V^\mu$ that still must be integrated over in a path integral,
for example.
The techniques for expanding a determinant in powers of
derivatives can be found in Ref.\cite{AITCHISON}; see also
the heat-kernel method in Ref.\cite{DONOGHUE92}.
The expansion of Eq.~(\ref{eq:Seff}) in a {\it renormalizable\/} model has
been discussed in Ref.\cite{PERRY86}.

We first focus on the nonderivative terms, which can be obtained
from Eq.~(\ref{eq:Seff}) by treating the fields as constants and
by expanding the logarithm in a power series in the fields.
Baryon number conservation implies that for the vector field,
only its derivatives can appear in the expansion.
Thus the nonderivative part of ${\cal L}_{\text{fd}}$
is an infinite polynomial in $\phi$;
for example, at the one-loop level,
\begin{eqnarray}
{\cal L}_{\text {fd}}[\phi]&=&i\int\!\! {{\rm d}^\tau k \over
        (2\pi)^4}       \,{\rm tr}\, \ln
      G^0(k)
       + i \sum_{n=1}^{\infty} {(-1)^{n}\over n} [g_{\rm s}\phi(x)]^n
       \int\!\! {{\rm d}^\tau k \over (2\pi)^4}\,
       {\rm tr}\,
         [G^0(k)]^n
               \ . \label{eq:cts}
\end{eqnarray}
Here we have regularized dimensionally to maintain Lorentz covariance and
baryon number conservation,
``${\rm tr}$'' denotes a trace over spin and isospin only, and
\begin{equation}
     G^0(k)
        ={1 \over \rlap/{\mkern-1mu k}-M+i\epsilon}
\end{equation}
is the free baryon propagator.
Beyond one loop there are additional terms in the coefficients,
including baryon counterterm contributions.

The polynomial in $\phi$ of Eq.~(\ref{eq:cts})
must be combined with the corresponding counterterms; in this way
the vacuum contributions are absorbed into the
renormalization of the scalar polynomial.
If one insists that the low-energy theorems be satisfied at tree
level in the meson fields,
the end result for the scalar potential should be of the form in
Eq.~(\ref{eq:NLag}), where the couplings are renormalized.
(Note that this potential can be expanded as a polynomial in $\phi$,
with {\it all\/} coefficients determined by $H_q$, $S_0$, and $d$.)
One never has to explicitly calculate any counterterms
or evaluate Eq.~(\ref{eq:cts}); when we write down the scalar potential,
the nucleon-loop effects have already been taken into
account.
Furthermore, although we have illustrated the renormalization by evaluating
nucleon loops only, any additional baryonic
degrees of freedom in the lagrangian would be treated analogously and the
final result will be the same.
Thus the phenomenological fitting of parameters accommodates a
general characterization of the vacuum response.

The renormalization of the derivative terms is analogous except
that we
do not have low-energy theorems to reduce the number of renormalized
coupling constants.
We note, however, that each additional derivative is accompanied by an
inverse power of a
typical scale in the problem, which is the nucleon mass here.
Experience with mean-field models of nuclei also suggest that the
derivatives
of the mean fields are small
(for example, $|\nabla\phi / \phi |
\lower0.6ex\vbox{\hbox{$\ \buildrel{\textstyle <}
         \over{\sim}\ $}} 100\,$MeV).
Thus if we assume
mean-field dominance, such that fluctuations around
the mean fields are small, and the naturalness of the
coefficients in the derivative expansion (see the discussion in Section~V),
we can truncate the derivative terms at some tractable order.
In this work, we will stop at the lowest order for the derivatives.
Thus we  have only a few  unknown renormalized constants (parameters),
which  are determined by fitting to experiment;
in our case, we will use finite-density observables.

At finite density, we work in the grand canonical ensemble through the
introduction of a chemical potential $\mu$\cite{TANGTHS}.
We consider only zero temperature in this work, which allows a
simplified discussion.
The relevant lagrangian density is now
\begin{equation}
  { \cal{L}}'(x,\mu)= {\cal{L}}(x) +\mu  \overline N\gamma_0N
                  \ .
\end{equation}
Here the effective action of ${ \cal{L}}'$ is associated with
the thermodynamic potential $\Omega$ of the system, instead of the
energy. The energy follows from
\begin{equation}
 E = \Omega + \mu B      \label{eq:En}         \ ,
\end{equation}
where
\begin{equation}
    B=-{\partial \Omega \over \partial \mu}  \label{eq:Bn}
\end{equation}
is the baryon number of the system.

Now we integrate out the baryon field as at zero density. The result
is the fermion determinant at finite density (or chemical potential),
$-i\, {\rm Tr}\ln K(\mu)$, to which
we can add and subtract the fermion determinant at $\mu=0$,
$-i\, {\rm Tr}\ln K(0)$.%
\footnote{We assume that $\mu=0$ still separates the positive-energy
levels from the Dirac sea.  This will be the case if the density
is not too high.}
The added term $-i\, {\rm Tr}\ln K(0)$ combines with the
counterterms exactly as described above
so that the renormalization goes through as before.
Note that it
contains the same dynamical scalar and vector fields as the fermion
determinant at $\mu$.
The remaining combination
\begin{equation}
   -i\, {\rm Tr}\ln K(\mu) + i\, {\rm Tr}\ln K(0)
\end{equation}
is an explicitly density-dependent piece (it vanishes for $\mu=0$),
which is finite if baryon counterterms are included in $K(\mu)$.
%which requires no renormalization \cite{FREEDMAN,MORLEY}.
(This combination is evaluated in the Hartree approximation in the next
section, for which the baryon counterterms are not needed.)
Once again the scalar potential in the form shown in
Eq.~(\ref{eq:NLag}) is left intact;
the only difference is that the scalar field now acquires a different
expectation value due to the presence of valence nucleons at finite density.

%%%%%%%%%%%%%%%%%%%%%%%%%%%%%%%%%%%%%%%%%%%%%%%%%%%%%%%%%%%%%%%%%%%%%%
%%%%%%%%%%%%%%%%%%%%%%%%%%%%%%%%%%%%%%%%%%%%%%%%%%%%%%%%%%%%%%%%%%%%%%

\section{Finite Nuclei and Nuclear Matter}

To perform a realistic calculation, we need a good starting approximation.
Since our focus here is on bulk nuclear properties and on single-particle
spectra, we assume that the mean meson fields dominate the dynamics, and
we expand the finite-density thermodynamic potential around the mean fields.
The lowest-order result (Hartree approximation)
is obtained by replacing all the meson fields
by their mean values, and this will be the starting point of any systematic
approximation for treating the fluctuations.

The thermodynamic potential for nuclei in the Hartree approximation
is given by
\begin{equation}
 \int\!{\rm d}x_0\,
      \Omega =i\, {\rm Tr}\ln \overline{K}(\mu) - i\, {\rm Tr}\ln
            \overline{K}(0)-\int {\rm d}^4x\, U_{\rm m}({\bf x})\ ,
             \label{eq:Omega}
\end{equation}
where the baryon kernel in coordinate space is now
\begin{equation}
  \langle x |  \overline{K}(\mu) | y \rangle  =\gamma_0
         [i\partial_0+\mu-h({\bf x})]\delta^{(4)}(x-y)
                         \ .
\end{equation}
The single particle hamiltonian $h$ is
\begin{equation}
  h({\bf x}) = -i\bbox{\alpha\cdot\nabla}
        +g_{\rm v}V_0({\bf x})+
                  \beta(M-g_{\rm s}\phi_0({\bf x}))\ ,
                     \label{eq:sph}
\end{equation}
with $\beta = \gamma_0$ and $\bbox{\alpha}=\gamma_0\bbox{\gamma}$,
and the static scalar and vector
mean fields are denoted by  $\phi_0({\bf x})$ and $V_0({\bf x})$.
The contribution from the meson fields is
\begin{eqnarray}
U_{\rm m}({\bf x}) &=&
          -{1\over 2}(\bbox{\nabla}\phi_0)^2
          -{1\over 4} m_{\rm s}^2 S_0^2 d^2
           \Big \{ \biggl(1-{\phi_0\over S_0}\biggr)^{4 / d}
                 \Big[{1 \over d}\ln \biggl(1-{\phi_0\over S_0}\biggr)
                  -{1\over 4}\Big ]+{1  \over 4}\Big \}
                   \nonumber   \\[6pt]
     & &  \null +{1\over 2}(\bbox{\nabla} V_0)^2+{1\over 2}
              \left(1+\eta {\phi_0\over S_0}\right)m_{\rm v}^2 V_0^2
                +{1\over 4!}\zeta (g_{\rm v} V_0)^4
                   \ .
\end{eqnarray}

Note that $\overline{K}(\mu)$ is diagonal in the single-particle basis
$\psi_{\alpha}({\bf x}) e^{i\omega x_0} $, where
$\psi_{\alpha}({\bf x})$ are the normalized
eigenfunctions of the Dirac equation with eigenvalues
$E_\alpha$ \cite{HOROWITZ,SW}:
\begin{equation}
 h \psi_{\alpha}({\bf x}) = E_{\alpha} \psi_{\alpha}({\bf x})\ , \
\ \ \ \
     \int {\rm d}^3x\,
\psi^{\dagger}_{\alpha}({\bf x})\psi_{\alpha}({\bf x})
     = 1 \ .  \label{eq:norm}
\end{equation}
{}From a path integral formulation, one can see that
the appropriate boundary condition
or  $i\epsilon$ prescription for evaluating the baryon kernel
is $\omega\rightarrow (1+i\epsilon)\omega$.

{}From Eq.~(\ref{eq:Omega}) one can now obtain,
after a Wick rotation,
\begin{eqnarray}
 \Omega &=&
       -\sum_{\alpha}\int {{\rm d}\omega \over 2\pi}\,
             [\ln (-i\omega +\mu-E_{\alpha}) - \ln (-i\omega
               -E_{\alpha})]
             -\int {\rm d}^3 x \, U_{\rm m}
                       \nonumber \\
          &=& -\sum_{\alpha} (\mu -E_{\alpha})
           [\theta(\mu -E_{\alpha})-\theta(-E_{\alpha})]
              -\int {\rm d}^3 x \, U_{\rm m}
                        \nonumber \\
          &\equiv& -\sum_{\alpha}^{\text{occ}}\, (\mu - E_{\alpha})
              -\int {\rm d}^3 x \, U_{\rm m}
                        \ . \label{eq:Om}
\end{eqnarray}
Here we have used
\begin{equation}
 \sum_{\alpha}\theta(-E_{\alpha})=\sum_{\alpha}\theta(E_{\alpha})
        =\sum_{\alpha}{1\over 2} \ ,
\end{equation}
which is valid when  $\mu=0$ separates the nucleon levels
from the antinucleon levels.
The summation superscript
``occ'' means that the sum runs only over occupied states in the Fermi sea.
Moreover, using Eqs.~(\ref{eq:En}) and (\ref{eq:Bn}), we find
\begin{eqnarray}
  B &=& \sum_{\alpha}\, [\theta(\mu -E_{\alpha})-\theta(-E_{\alpha})]
         =  \sum_{\alpha}^{\text{occ}} \, 1 \ ,\\
  E &=& \sum_{\alpha}^{\text{occ}} E_{\alpha}
        -\int {\rm d}^3 x \, U_{\rm m} \ . \label{eq:Eeqn}
\end{eqnarray}
We emphasize that the final sum over only occupied (valence) states
is not the result of a {\it vacuum\/} subtraction,
as the term with $\mu=0$ still contains the background fields,
which must be determined self-consistently.
The true vacuum subtraction was performed earlier when we derived the
renormalized $U_{\rm m}$.

The equations for the mean fields are obtained from
extremizing the energy functional
with respect to $\phi_0({\bf x})$ and $V_0({\bf x})$.
{}From Eqs.~(\ref{eq:sph}) and (\ref{eq:norm}) one finds
\begin{eqnarray}
        {\delta E_\alpha\over\delta\phi_0({\bf x})}
         & = & {\delta\over\delta\phi_0({\bf x})}
          \int\!{\rm d}^3y\, \psi^\dagger_\alpha({{\bf y}}) h({{\bf y}})
                   \psi_\alpha({{\bf y}})
        \nonumber \\
          & = & \psi^\dagger_\alpha({{\bf x}})
          {\partial h\over \partial\phi_0} \psi_\alpha({{\bf x}}) +
            E_\alpha {\delta\over\delta\phi_0({\bf x})}
          \int\!{\rm d}^3y\, \psi^\dagger_\alpha({{\bf y}})
                   \psi_\alpha({{\bf y}})
        \nonumber \\
         & = & \psi^\dagger_\alpha({{\bf x}}) {\partial h\over
\partial\phi_0}
         \psi_\alpha({{\bf x}})
           \ ,
        \label{eq:inter}
\end{eqnarray}
and a similar expression for the variation with respect to $V_0$;
evaluating the derivatives yields
\begin{eqnarray}
 {\delta \over \delta \phi_0 ({\bf x})} \sum_{\alpha}^{\text{occ}}
E_{\alpha}
        &=& -g_{\rm s} \sum_{\alpha}^{\text{occ}}
              \overline \psi_{\alpha}({\bf x})\psi_{\alpha}({\bf x})
                         \ , \\
 {\delta \over \delta V_0 ({\bf x})} \sum_{\alpha}^{\text{occ}}
E_{\alpha}
        &=& g_{\rm v} \sum_{\alpha}^{\text{occ}}
               \psi^\dagger_{\alpha}({\bf x})\psi_{\alpha}({\bf x})
\ .
\end{eqnarray}
Upon applying these results to Eq.~(\ref{eq:Eeqn}),
one obtains the  mean-field equations:
\begin{eqnarray}
   -{\bbox{\nabla}}^2 \phi_0 + m_{\rm s}^2 \phi_0  & = &
     g_{\rm s} \sum_{\alpha}^{\text{occ}}
              \overline \psi_{\alpha}({\bf x})\psi_{\alpha}({\bf x})
                \nonumber
      \\ & & \null      + m_{\rm s}^2 \phi_0 +
            m_{\rm s}^2 S_0 \biggl(1-{\phi_0\over S_0}\biggr)^{(4/ d)
-1}
            \ln \biggl(1-{\phi_0\over S_0}\biggr) +
            {\eta \over 2 S_0} m_{\rm v}^2 V_0^2   \ , \\
   -{\bbox{\nabla}}^2 V_0 + m_{\rm v}^2 V_0  & = &
        g_{\rm v} \sum_{\alpha}^{\text{occ}}
               \psi^\dagger_{\alpha}({\bf x})\psi_{\alpha}({\bf x})
             \nonumber
      \\ & &  \null     - \eta {\phi_0\over S_0} m_{\rm v}^2 V_0
           - {1\over 6} \zeta g_{\rm v}^4 V_0^3  \ .  \label{eq:Vfeqn}
\end{eqnarray}
Note that we have added an explicit mass term to each side of the
scalar field equation to put it in a form that can be solved
with conventional numerical techniques \cite{HOROWITZ}.
The vector mean-field equation is actually a constraint since the
time component of the vector field is not a dynamical degree of freedom.
(See below for further comments in the case of nuclear matter.)

This lowest-order result (Hartree approximation)
is similar to that obtained from conventional derivations
of relativistic mean-field models in which one-loop vacuum corrections
are simply neglected.  We emphasize, however, that we are not merely
presenting another mean-field model; the vacuum effects {\it are\/}
incorporated and systematic improvement is possible (in principle).
Rather, we consider our procedure a {\it justification\/}
for the phenomenologically successful mean-field approach.

The energy density for uniform nuclear matter in the Hartree
approximation can be obtained from the preceding results
by observing that the single-particle energy eigenvalue becomes
\begin{equation}
 E({\bf k})=g_{\rm v} V_0 +\sqrt{{\bf k}^2+{M^*}^2} \ ,
\end{equation}
where $M^*=M-g_{\rm s}\phi_0$, and $\phi_0$ and $V_0$ are now constant
mean fields. The energy density ${\cal E}$ becomes
\begin{eqnarray}
     {\cal E}[M^\ast,\rho_{{\scriptscriptstyle\rm B}}]
        &=& {1\over 4} m_{\rm s}^2 S_0^2 d^2
           \Bigl\{ \Bigl(1-{\phi_0\over S_0}\Bigr)^{4/d}
                 \Bigl[{1 \over d}\ln \Bigl(1-{\phi_0\over S_0}\Bigr)
                    -{1\over 4}\Bigr]
                  + {1\over 4} \Bigr\}
           +g_{\rm v}\rho_{\scriptscriptstyle\rm B}
            V_0 - {1\over 4!}\zeta
             (g_{\rm v} V_0)^4
                   \nonumber \\[4pt]
     & &         - {1 \over 2}\Bigl(1+\eta {\phi_0\over S_0}\Bigr)
                   \, m_{\rm v}^2
             V_0^2
     + {\gamma\over (2\pi)^3}\! \int^{k_{\scriptscriptstyle\rm F}}
       {\kern-.1em}{\rm d}^3{\kern-.1em}{k} \,
           \sqrt{{\bf k}^2+{M^*}^2}
          \, ,         \label{eq:endens}
\end{eqnarray}
where $k_{\scriptscriptstyle\rm F}$ is the Fermi momentum
defined by $\mu =g_{\rm v}V_0+\sqrt{k_{\rm \scriptscriptstyle F}^2+{M^*}^2}$,
and
$\rho_{\scriptscriptstyle\rm B}=
\gamma k_{\scriptscriptstyle\rm F}^3/(6\pi^2)$
is the baryon density.
The spin-isospin degeneracy $\gamma = 4$
for nuclear matter and $\gamma = 2$ for
neutron matter.

The equation that determines $V_{ 0}$
can be obtained either from the Euler--Lagrange equations or by using
Dirac's procedure \cite{DIRAC}, with the result
\begin{equation}
g_{\rm v}\rho_{\scriptscriptstyle\rm B}=
      \Bigl(1+\eta {\phi_0\over S_0}\Bigr)\,
           m_{\rm v}^2 V_{ 0}
        + {1\over 6}\zeta g_{\rm v}^4 V_{ 0}^3
           \ . \label{eq:vcnstrnt}
\end{equation}
This equation can also be obtained
from Eq.~(\ref{eq:endens}) by setting
$(\partial {\cal E}/\partial V_{ 0})_{
\rho_{\scriptscriptstyle\rm B}, M^*}=0\ $.
Note, however, that this is not a minimization condition for ${\cal E}$.
In fact, the $V_{ 0}$ obtained from Eq.~(\ref{eq:vcnstrnt})
corresponds to a local maximum of the energy density.
Equation~(\ref{eq:vcnstrnt}), like Eq.~(\ref{eq:Vfeqn}),
is  a constraint equation for
$V_{ 0}$, which is not a dynamical variable.

\begin{figure}[tbhp]
\centerline{%
\vbox to 3.5in{\vss
   \hbox to 3.3in{\includegraphics{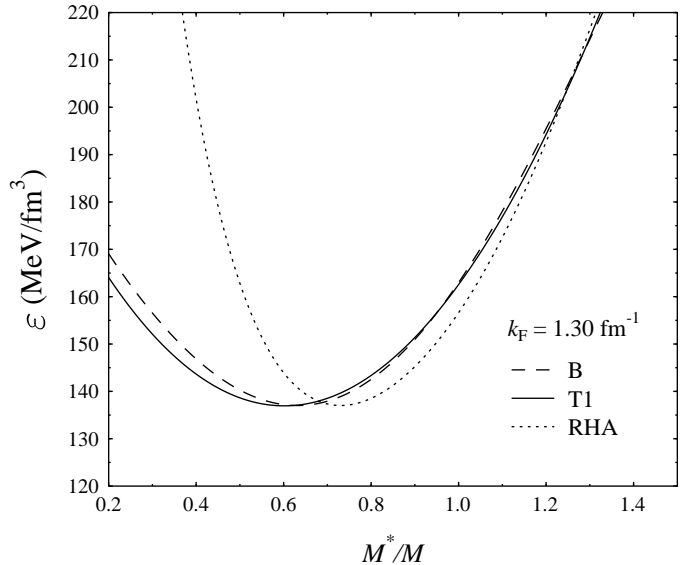}\hss}}
}
\caption{\capcrunch{%
Finite-density
effective potential ${\cal E}$ from Eq.~(\protect\ref{eq:endens}),
  plotted as a function of $M^\ast$ (solid line).  $V_{ 0}$ is
eliminated for each $M^\ast$ using Eq.~(\protect\ref{eq:vcnstrnt}).
Parameter set T1 is used and
$k_{{\scriptscriptstyle\rm F}} = 1.30\mbox{\,fm}^{-1}$.
Results for other parameter sets and other densities are
qualitatively similar.
Also shown are the analogous potentials for the Walecka model
RHA \protect\cite{WALECKA74} (dotted line) and the
nonlinear parameter set B from Ref.~\protect\cite{FS} (dashed line).}
}
\label{fig:one}

\end{figure}

The energy density at a given baryon density is found
by using Eq.~(\ref{eq:vcnstrnt}) to eliminate
$V_{ 0}$ from ${\cal E}$ in Eq.~(\ref{eq:endens})
and then by minimizing the resulting finite-density
effective potential with respect
to $M^\ast$.
The effective potential at fixed baryon density is shown in Fig.~\ref{fig:one}.
Notice that in contrast to the conventional one-loop approximation
(relativistic Hartree approximation or RHA \cite{WALECKA74,SW}) in
renormalizable models, the finite-density
effective potential of our truncated model is meaningful
only when $|g_{\rm s}\phi_0 |$ is sufficiently
small that higher-order terms can be neglected.
Similar considerations apply to the solutions of Eq.~(\ref{eq:vcnstrnt}).
(See Section~V for further discussion.)
Parameters can be chosen so that nuclear matter exhibits saturation at
the empirical point;
one approach to determining the parameters is discussed in the next
section.

%%%%%%%%%%%%%%%%%%%%%%%%%%%%%%%%%%%%%%%%%%%%%%%%%%%%%%%%%%%%%%%%%%%%%%
%%%%%%%%%%%%%%%%%%%%%%%%%%%%%%%%%%%%%%%%%%%%%%%%%%%%%%%%%%%%%%%%%%%%%%

\section{Results}

To test the utility of the model,
we must see if it can successfully describe finite nuclei \cite{FS}.
The basic features we seek to reproduce are the nuclear charge densities
(including the observed flatness in heavy nuclei), the characteristics
of the single-particle spectrum, and the bulk binding-energy
systematics.
Relativistic mean-field models unconstrained by QCD symmetries
have been successful in reproducing these properties for nuclei
across the periodic table.

The Hartree equations for finite nuclei in our model were given
in Section III, but only isoscalar mesons were discussed.
To make realistic comparisons to experiment, we must include
the $\rho$ and the Coulomb interactions.
Here we simply introduce
the $\rho$ and the photon  as in Ref.~\cite{FS}, except that we also
include a coupling between the $\rho$ and the scalar $\phi$, exactly as for the
$\omega$ [see Eq.~(\ref{eq:endens})].
A more complete treatment of the isovector mesons
will be presented elsewhere\cite{FST}.

We take the nucleon, $\omega$, and $\rho$ masses as given
by their experimental values:
$M=939\,$MeV, $m_{\rm v} = 783\,$MeV, and $m_{\rho}=770\,$MeV.
We then fit the rest of the parameters
($g_{\rm s}$, $g_{\rm v}$,
$g_{\rho}$, $\eta$, $\zeta$, $m_{\rm s}$, $S_0$, and $d$)
to the binding energies, the charge radii,
and the spin-orbit splittings of the least-bound proton and
neutron in $^{16}$O, $^{40}$Ca, and $^{208}$Pb, as well as to
the charge density of $^{16}$O at $r \approx 1\,$fm.
An optimization process similar to that of Ref.~\cite{LosAlamos} is used.
Here we are principally interested in showing that a good fit to properties
of finite nuclei {\it can\/} be achieved;
Table~\ref{tab:one} lists three such parameter sets (T1, T2, and T3).
In set T1, $d$ is an optimization parameter, while it is fixed
(arbitrarily) in sets T2 and T3 to illustrate the range of possible $d$.
In a future paper, we will study in more detail the regions of the parameter
space that produce a reasonable fit and examine which conditions are important
in determining individual parameters.

\begin{table}[tbh]
\caption{Parameter sets
from fits to finite nuclei.
The vector masses are $m_{\rm v} = 783\,$MeV and $m_\rho = 770\,$MeV;
the nucleon mass is $M = 939\,$MeV.
Values for $S_0$, the scalar mass $m_{\rm s}$, and
$H_{\rm q}^{1/4}$ are in MeV.
Note that $m_{\rm s}^2 = 4 H_{\rm q}/(d^2 S_0^2)$.
}
\smallskip
\begin{tabular}[tbh]{cccccccccc}
 Set & $g_{\rm s}^2$ & $m_{\rm s}$ & $g_{\rm v}^2$ & $g_{\rho}^2$ &
$S_0$
                  & $\zeta$ & $\eta$ & $d$ & $H_{\rm q}^{1/4}$ \\
 \hline
 T1  &   99.3   &  509. & 154.5  & 70.2
    &  90.6  &  0.0402  &  $-0.496$ &  2.70
    & 250. \\
 T2  &   96.3   &  529. & 138.0  & 69.6
    &  95.6  &  0.0342  &  $-0.701$ &  2.20
    & 236. \\
 T3  &  109.5   &  508. & 178.6  & 67.2
    &  89.8  &  0.0346  &  $-0.160$ &  3.50
    & 283. \\
\end{tabular}
\label{tab:one}
\end{table}

\narrowtext
\begin{table}[tbh]
\caption{Binding-energy systematics for the model proposed here
(sets T1, T2, and T3), for model B from Ref.~\protect\cite{FS},
and for the point-coupling
(PC) model of Ref.~\protect\cite{LosAlamos}.
Binding energies per nucleon are given in MeV.}
\begin{tabular}[tbh]{cccc}
 Model    & $^{16}$O   & $^{40}$Ca  &  $^{208}$Pb \\ \hline
 T1    &  7.99  &  8.61  &  7.91  \\
 T2    &  7.94  &  8.55  &  7.89  \\
 T3    &  7.95  &  8.53  &  7.91  \\
 B     &  7.82  &  8.35  &  7.62  \\
 PC    &  7.97  &  8.58  &  7.87  \\
exp't  &  7.98  &  8.55  &  7.87  \\
\end{tabular}
\label{tab:two}
\end{table}

\narrowtext
\begin{table}[tbh]
\caption{Rms charge radii (in fm)
for  the model proposed here
(sets T1, T2, and T3), for model B from Ref.~\protect\cite{FS},
and for the point-coupling
(PC) model of Ref.~\protect\cite{LosAlamos}.}
\begin{tabular}[tbh]{cccc}
 Model    & $^{16}$O   & $^{40}$Ca  &  $^{208}$Pb \\ \hline
 T1    &  2.73  &  3.47  &  5.56  \\
 T2    &  2.72  &  3.47  &  5.56  \\
 T3    &  2.72  &  3.48  &  5.57  \\
 B    &  2.74  &  3.48  &  5.56  \\
 PC   &  2.73  &  3.45  &  5.51  \\
exp't &  2.74  &  3.47  &  5.50  \\
\end{tabular}
\label{tab:three}
\end{table}

\begin{figure}[tbhp]
\centerline{%
\vbox to 3.5in{\vss
   \hbox to 3.3in{\includegraphics{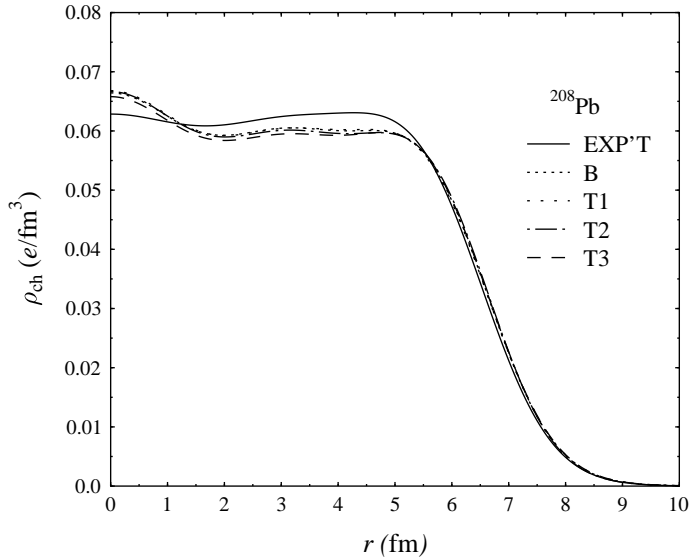}\hss}}
}
\caption{\capcrunch{%
Charge density of $^{208}$Pb.
The solid line is taken from experiment \protect\cite{DEVRIES87}.
Charge densities are shown for
a successful mean-field model (model B from
Ref.~\protect\cite{FS}) and for the three parameter sets
from Table~I.}
}
\label{fig:two}
\end{figure}

\begin{figure}[tbhp]
\centerline{%
\vbox to 5.in{\vss
   \hbox to 3.3in{\includegraphics{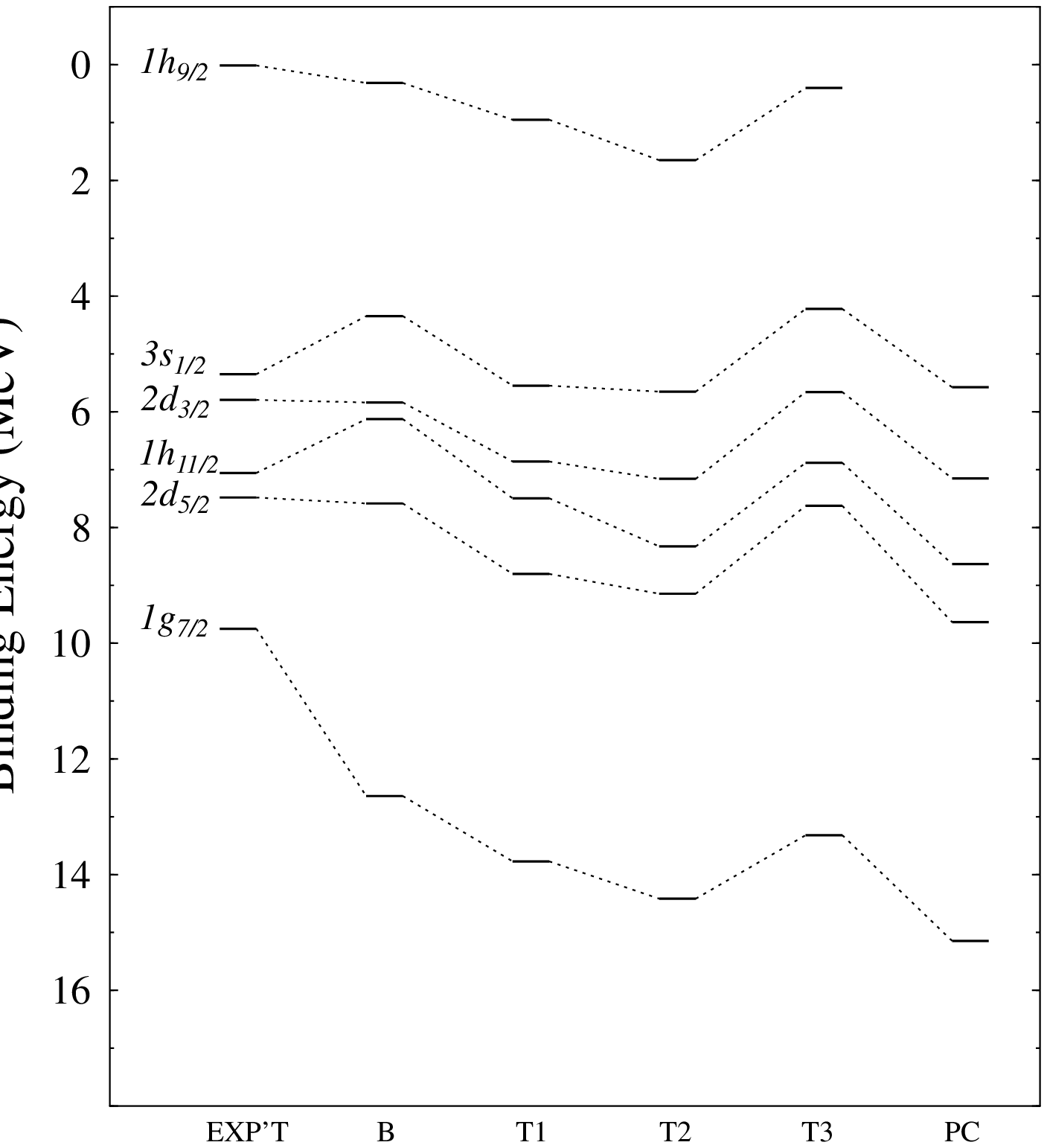}\hss}}
}
\caption{\capcrunch{%
Predicted proton single-particle spectra for $^{208}$Pb using
the parameter sets from Table~I.
Only the least-bound major shell is shown.
The leftmost values are from experiment,  model B
is a successful mean-field model from
Ref.~\protect\cite{FS}, and model PC is the point-coupling
model of Ref.~\protect\cite{LosAlamos}.
Note that the $1h_{9/2}$ level is an {\it unoccupied\/} state.}
}
\label{fig:three}

\end{figure}

\begin{figure}[tbhp]
\centerline{%
\vbox to 3.5in{\vss
   \hbox to 3.3in{\includegraphics{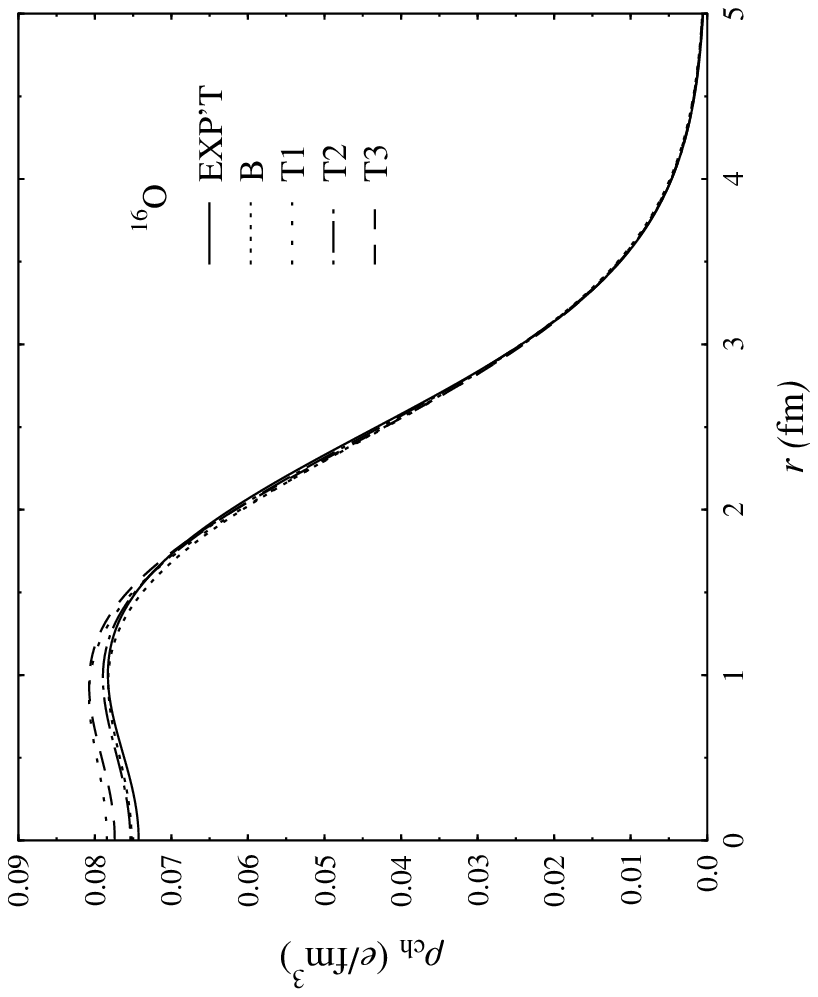}\hss}}
}
\caption{\capcrunch{%
Charge density of $^{16}$O.
The solid line is taken from experiment \protect\cite{DEVRIES87}.
Charge densities are shown for
a successful mean-field model (model B from
Ref.~\protect\cite{FS}) and for the three parameter sets
from Table I.}
}
\label{fig:four}
\end{figure}

\begin{figure}[tbhp]
\centerline{%
\vbox to 3.5in{\vss
   \hbox to 3.3in{\includegraphics{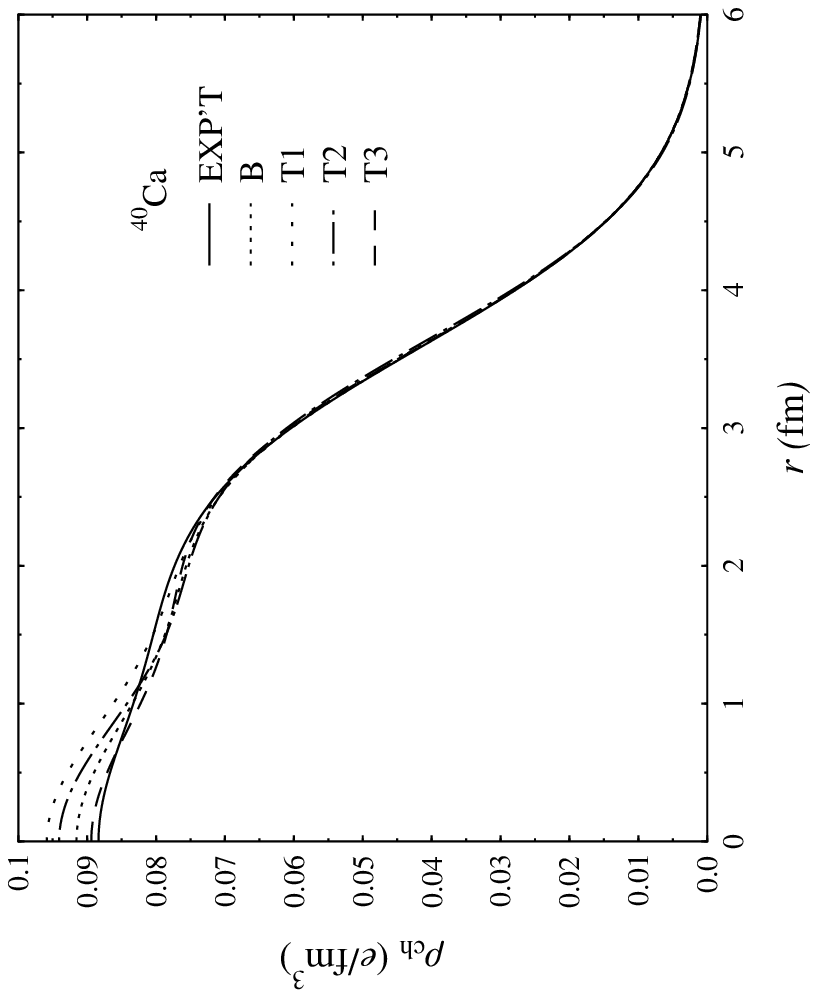}\hss}}
}
\caption{\capcrunch{%
Charge density of $^{40}$Ca.
The solid line is taken from experiment \protect\cite{DEVRIES87}.
Charge densities are shown for
a successful mean-field model (model B from
Ref.~\protect\cite{FS}) and for the three parameter sets
from Table I.}
}
\label{fig:five}

\end{figure}

We have calculated $^{16}$O, $^{40}$Ca, and $^{208}$Pb for these
parameter sets and
for a representative mean-field model (set B from Ref.~\cite{FS}).
Bulk binding-energy systematics are summarized in Table~\ref{tab:two}
and rms charge radii are summarized in Table~\ref{tab:three}.
For comparison, we also include results from the point-coupling model
of Ref.~\cite{LosAlamos}.
The binding energies include center-of-mass corrections as in
Ref.~\cite{REINHARD89}.
We show charge densities and single-particle levels for
$^{208}$Pb in Figs.~\ref{fig:two} and \ref{fig:three},
and charge densities for $^{16}$O and $^{40}$Ca in Figs.~\ref{fig:four}
and \ref{fig:five}.
The charge densities are determined from point-proton densities following
the conventional procedure \cite{HOROWITZ}, which folds them with
a phenomenological proton form factor.
Form factors generated within the model itself, originating from vector
dominance physics, will be considered elsewhere.

The fits to nuclear charge radii, binding energies, and spin-orbit splittings
are quite good.
The only  deficiencies in the sets illustrated here are some
small deviations from
experiment in the  charge densities.
Changes in the optimization procedure can improve the agreement of the
charge densities at the cost of worsening slightly the agreement with
empirical binding energies.

A good reproduction of the spin-orbit force in finite nuclei necessarily
leads to large scalar and vector mean fields in the interiors of the nuclei
or in nuclear matter.
In particular, as discussed many times (recently by Bodmer \cite{BODMER91}),
vector and scalar fields of roughly 250--300\,MeV are needed to reproduce
the observed spin-orbit splittings in the least-bound levels (and also
the deformations in light, axially symmetric nuclei \cite{FPW}).
While these fields are large on the scale of the nuclear binding energies,
$| g_{\rm v}V_0 |/M$ and
$| g_{\rm s}\phi_0 |/M$ and their gradients in finite nuclei
are relatively small; thus, these remain useful
expansion parameters.
This justifies our truncation of the energy density at small powers of the
meson fields.
While it is possible in principle to add additional monomials in the fields
(with undetermined parameters), the quality of the present fit makes it
unlikely
that there is much to be gained by this.

The scale dimension $d$ of the light scalar field  was found to
be about 2.7 when $d$ was included in the optimization.
Note that the canonical dimension would have $d=1$.
Changes in the optimization procedure or a relaxation in the goals of
the fit allow for a considerable range in $d$
(sets T2 and T3 are examples), but it does not seem possible
to find a reasonable parameter set with $d<2$.
{\em Thus the introduction of an anomalous dimension for the light scalar
degree of freedom is an essential feature for the phenomenological
success of our model}.

\begin{table}[tbh]
\caption{Nuclear matter saturation properties for the model proposed
here
(sets T1, T2, and T3), for model B from Ref.~\protect\cite{FS},
and for the point-coupling
(PC) model of Ref.~\protect\cite{LosAlamos}.
Values are given for the binding energy per nucleon (in MeV),
the Fermi momentum $k_{{\scriptscriptstyle\rm F}}$ (in fm$^{-1}$), the
compressibility $K$ (in MeV), the bulk symmetry energy coefficient
$a_4$ (in MeV), $M^\ast /M$,
and $g_{\rm v} V_0$ (in MeV) at equilibrium.
}
\smallskip
\begin{tabular}[tbh]{cccccccc}
 Model & $E/B-M$ & $k_{{\scriptscriptstyle\rm F}}$
            & $K$ & $a_4$ & $M^\ast/M$ & $g_{\rm v} V_0$  \\
 \hline
 T1   & 16.2 & 1.30 & 194. & 39. & 0.60 & 302. \\
 T2   & 16.3 & 1.29 & 240. & 40. & 0.61 & 298. \\
 T3   & 16.1 & 1.29 & 244. & 34. & 0.61 & 297. \\
 B   & 15.8 & 1.30 & 220. &  35. & 0.63 & 277. \\
 PC  & 16.1 & 1.30 & 264. &      & 0.58 & 322. \\
\end{tabular}
\label{tab:oneb}
\end{table}

Experience with a broad class of relativistic mean-field models
shows that models that successfully reproduce bulk properties of finite
nuclei share characteristic properties in infinite nuclear matter \cite{FS}.
These properties
are the equilibrium binding energy and density, the compressibility $K$,
and the value of $M^\ast/M$ at equilibrium.
One further condition, that the light scalar mass $m_{\rm s} \approx 500\,$MeV,
is needed to ensure reasonably smooth charge densities
and good surface-energy systematics.
If we calculate nuclear matter with the parameter sets in
Table~\ref{tab:one}, we find good agreement with values found
in investigations with unconstrained mean-field models
 (see Table~\ref{tab:oneb}).
In particular, the saturation density corresponds to a Fermi momentum
of about 1.3\,fm$^{-1}$,
and the binding energy per nucleon at saturation is about 16\,MeV.
The compressibility is less well determined (190--250\,MeV).
The nucleon effective mass
$M^*/M \approx 0.60$ and the scalar mass $m_{\rm s}$ is just over
500\,MeV.
We emphasize that these values are obtained after a fit to finite nuclei only.

%%%%%%%%%%%%%%%%%%%%%%%%%%%%%%%%%%%%%%%%%%%%%%%%%%%%%%%%%%%%%%%%%%%%%%
%%%%%%%%%%%%%%%%%%%%%%%%%%%%%%%%%%%%%%%%%%%%%%%%%%%%%%%%%%%%%%%%%%%%%%

\section{Discussion}

We can relate the phenomenological success of the model proposed here
to the characteristics of successful relativistic mean-field models
of finite nuclei.
A key feature is the logarithmic
potential for the scalar field, which allows for relatively
weak nonlinearities and the dominance of the cubic and quartic scalar terms,
with the values of the scaling dimension $d$ used here.
In contrast, chiral models with a Mexican hat potential have large cubic
and quartic terms, which preclude a good fit to bulk nuclear properties
\cite{FS}.
Bodmer \cite{BODMER91} has shown that nuclear matter properties that lead
to good predictions for finite nuclei can be achieved if one adds to (small)
cubic and quartic scalar terms a term that is quartic in the vector field
(here with coupling $\zeta$).
Thus our model has all of the  ingredients needed to allow a good fit
through optimization.
In addition,  adjustments can be made through the
scalar-vector coupling $\eta$.

Note that the scalar-vector coupling and the quartic vector
self-coupling can be used to define an
effective, density-dependent mass of the vector meson at the mean-field level.
For example, one can use
the second derivative of the lagrangian with respect to the vector field.
For the model parameters in Table~\ref{tab:one}, the two contributions
largely cancel, so that the vector effective mass $m_{\rm v}^\ast$
is essentially independent of density.
This is in contrast to the universal scaling hypothesis of Brown and Rho
\cite{BROWN91},
which predicts  $m_{\rm v}^\ast/m_{\rm v} = M^\ast/M$.

We have excluded many terms from our model: higher-order
polynomials in the vector fields and mixed scalar-vector terms,
non-Yukawa couplings to the nucleon, derivative terms, and so on.
In retrospect, were we justified in neglecting them?
An analysis of mean-field models \cite{BODMER91,FST}
implies that one can identify
dimensionless ratios that can be used to set the scale of individual
contributions to the energy.
For example,
one can rewrite the scaled energy density of nuclear matter,
${\cal E}/M^4$, in terms of the dimensionless ratios
$g_{\rm v}V_0 /M$ and $g_{\rm s}\phi_0 /M$,
which then become our finite-density expansion parameters.

Moreover, an important assumption in applying effective field theories,
such as chiral perturbation theory, is that the coefficients of terms in the
lagrangian are ``natural,'' {\it i.e.,\/} of order unity,
when written in appropriate dimensionless units.
This assumption makes the organization of terms through a
power-counting scheme useful, because one can systematically
truncate the expansion when working to a desired accuracy.
We have proposed an analogous concept of naturalness for the
finite-density problem,
which will justify the neglect of higher derivatives and
powers of the fields when applying Eq.~(\ref{eq:NLag}) to nuclei.
For example, if one expresses the nuclear matter energy density in terms of
the scaled field variables written above, one finds that the ratios
\begin{equation}
    {m_{\rm v}^2 \over 2 g^2_{\rm v} M^2} \ , \quad
    {m_{\rm s}^2 \over 2 g^2_{\rm s} M^2} \ , \quad
    {\zeta \over 8}\ , \quad
    {\eta m_{\rm v}^2 \over 2 g_{\rm s} g_{\rm v}^2 S_0 M} \ ,
       \label{eq:scaledstuff}
\end{equation}
should all be of roughly equal size for our expansion to be ``natural.''
One can verify that the values in sets T1, T2, and T3 satisfy this condition.

To examine the size of the scalar self-interactions, one expands the
logarithmic potential in Eq.~(\ref{eq:endens}) with the result
\begin{eqnarray}
    {1\over 4} m_{\rm s}^2 S_0^2 d^2
           \Bigl\{ \Bigl(1 &-&{\phi_0\over S_0}\Bigr)^{4/d}
                 \Bigl[{1 \over d}\ln \Bigl(1-{\phi_0\over S_0}\Bigr)
                    -{1\over 4}\Bigr]
                  + {1\over 4} \Bigr\}\Big/ M^4    \nonumber \\[4pt]
 & = & \Bigl[ {m_{\rm s}^2 \over 2! g_{\rm s}^2 M^2} \Bigr]
             {\widetilde \Phi}^2
    + \Bigl[ {1\over 3! M g_{\rm s}^3}\ {(3d - 8) m_{\rm s}^2\over d S_0}
		\Bigr] {\widetilde \Phi}^3
     + \Bigl[ {1\over 4! g_{\rm s}^4}\ {(11d^2 - 48d + 48) m_{\rm s}^2\over
             (d S_0)^2} \Bigr] {\widetilde \Phi}^4
\nonumber \\[4pt]
  & &   + \Bigl[ {M\over 5! g_{\rm s}^5}\ {2(25d^3 - 140d^2 + 240d - 128)
           m_{\rm s}^2\over (d S_0)^3} \Bigr] {\widetilde \Phi}^5
\nonumber \\[4pt]
 & & + \Bigl[ {M^2\over 6! g_{\rm s}^6}\ {2(137d^4 - 900d^3 + 2040d^2
          -1920d + 640) m_{\rm s}^2 \over (d S_0)^4} \Bigr]
           {\widetilde \Phi}^6 + \cdots  \ , \label{eq:logexp}
\end{eqnarray}
where ${\widetilde \Phi} \equiv g_{\rm s} \phi_0 /M$.
The coefficients in square brackets give the combinations that should be
compared to those in Eq.~(\ref{eq:scaledstuff}), and one can verify that
these are also natural for parameter sets T1, T2, and T3.
It is interesting that for set T1, in which $d$ is an optimized parameter,
the scaled coefficients are extremely small due to nearly complete
cancellations among terms in the polynomials in $d$.
(Further discussion of these issues is given in Ref.~\cite{FST}.)

Further support for the naturalness assumption comes from extending the
model to include $\phi^2 V_\mu V^\mu$ and $(V_\mu V^\mu)^3$ terms and then
repeating the optimization.
The new fit is very close to the fit obtained without these terms.
Furthermore, contributions to the energy from the new terms are
less than 10\% of those from the old terms at nuclear
matter density, and the old coefficients change only slightly in the
new fit \cite{FST}.
Thus contributions from the higher-order terms can be absorbed into
slight adjustments of the coefficients in Eq.~(\ref{eq:NLag}).

The astute reader will note that if our naturalness assumption is
justified, we could construct a variation of our model
without the constraints of the low-energy theorems of broken scale invariance.
Indeed, at nuclear matter density,
numerics alone would let us truncate the scalar potential, and the same
arguments about renormalization apply, so that vacuum effects are still
built in.
This explains the success of previous relativistic mean-field models of nuclear
structure and illustrates the power of the assumption of
naturalness.
Here we note that the scalar potential constrained by the low-energy theorems
actually provides some justification for naturalness, which we simply
assume is valid for higher-order and derivative terms.
Thus one should be cautious in drawing strong conclusions about the role of
broken scale invariance when applied in models that are restricted to
moderate nuclear density.

How widely can our model be applied?
A prime motivation for developing relativistic models of nuclei and
nuclear matter is to extrapolate to extremes of density and
temperature \cite{WALECKA74}.
Such conditions can be reached experimentally in relativistic heavy-ion
collisions.
One hopes that the calibration of such models to observables at
ordinary nuclear densities and zero temperature, in conjunction with
constraints from QCD symmetries, will permit reliable extrapolations.

Unfortunately, our framework of mean-field dominance, naturalness, and the
truncation at small powers of the fields and their derivatives,
which limits the number of parameters at
ordinary nuclear densities, is bound to break down as the density increases.
With increasing density, we will find increasing mean fields and
expansion parameters that are no longer small.
Thus we become increasingly less justified in ignoring the effects of
higher-order terms, and the calibration at nuclear matter density becomes
less and less of a constraint.
The limits of reliable extrapolation are not clear, but one should
certainly be cautious in applying models like ours much above nuclear
matter density.
Nevertheless, the utility of an accurate relativistic mean-field model for
nuclear structure and reactions, which is compatible with the low-energy
behavior of QCD, should be obvious.

We close our discussion with some interesting observations.
{}From Table~\ref{tab:one},
one sees that $S_0$, the vacuum expectation value of the
light scalar field $S$, is  close to the experimental value of
$f_\pi$ (93\,MeV).
Furthermore, the scalar coupling constant $g_{\rm s}$ is close to
$g_\pi/g_A$.
If we forget for the moment complications
from requiring terms to be scale invariant,
it is tempting to say that the model has a preference for the nucleon
mass to be generated entirely from the vacuum expectation value of $S$.
That is, if the only nucleon coupling to scalar fields is $g_{\rm
s}\overline NN S$, then we recover the empirical nucleon mass and the
Goldberger--Treiman relation from the fit values of the other parameters.
This scenario is also consistent with Miransky's model, in which the
light scalar (quarkonium) is associated with the quark condensate, and
with QCD sum rules, which associate the nucleon mass predominantly with
the quark condensate.
It is premature to do more than to point out these results, but the
coincidence of numbers certainly merits further investigation.

\section{Summary}

In summary, we have introduced a new model for nuclear matter and finite
nuclei that realizes QCD symmetries at the hadronic level.
In particular, the model incorporates chiral symmetry, broken scale
invariance, and the phenomenology of vector dominance.
An important feature is the light scalar degree of freedom, which
is given an anomalous scale dimension.
The renormalized scalar potential is constrained by the low-energy
theorems of broken scale invariance.
Vacuum loop effects are absorbed into the renormalized parameters,
which are determined by fits
to hadron masses and finite-density observables.

The truncation of the model lagrangian is based on mean-field
dominance and the identification of expansion parameters that
are reasonably small at nuclear matter densities.
Due to the characteristics of the constrained scalar potential,
we adopt a ``naturalness'' assumption, which justifies the truncation.
The parameters of the truncated model are identified by an optimization
procedure designed to reproduce bulk properties of finite nuclei.
Good fits are obtained, which also lead to very reasonable nuclear
matter properties.
The scale dimension of the scalar field comes out greater than two,
but is not tightly constrained by the fit.

It is important to emphasize what we have learned about the relationship
between effective (hadronic) theories of QCD and successful relativistic
mean-field phenomenology.
The vacuum dynamics of QCD is  constrained by the trace anomaly and the
consequent low-energy theorems of QCD.
At the level of hadronic fields, this physics manifests itself in the
scalar-isoscalar sector of the theory.
We have proposed that this sector can be divided into a low-mass part that
is adequately described by a scalar meson with anomalous dimension and a
high-mass part that is ``integrated out,'' leading to various couplings
among the remaining fields.
We believe this latter characterization is quite general and independent
of the details of the high-mass part of the scalar sector.
Nevertheless, whereas the realization of the Goldstone boson dynamics is
well known ({\em i.e.}, chiral perturbation theory), as is the dynamics of
the vector sector ({\em i.e.}, vector-meson dominance), little is known
about the precise form and magnitudes of the nonlinear couplings
originating from the scalar degrees of freedom.
We find that our primary source of information on this dynamics comes from
{\em nuclear structure physics}, which provides strong constraints on this
sector of the theory.

In subsequent work, we will further explore the parameter space that leads
to good fits to nuclear properties and identify the observables that
constrain individual terms.
We will also investigate the chiral properties of the model and
study the implications of vector dominance for nuclear observables.
Work to extend the model beyond the one-baryon-loop
level in a manner consistent with
conservation laws and Ward identities is in progress.
Finally, we will continue the development of the naturalness concept for
finite-density systems.

\acknowledgments

We are pleased to thank P. J. Ellis, S.~V. Gardner, C.~J. Horowitz,
D.~B. Leinweber, V.~A. Miransky,
R.~J. Perry, S.~Rudaz, and J.~Rusnak for useful comments.
This work was supported in part by the National Science
Foundation
under Grant Nos.~PHY-9203145, PHY-9258270, PHY-9207889, and
PHY-9102922, and the Sloan Foundation and
by the U.S. Department of Energy under
contract No.~DE--FG02--87ER40365.

\end{document}